\documentclass[%
 reprint,
 amsmath,amssymb,
 aps,
]{revtex4-2}

\usepackage{graphicx}
\usepackage{dcolumn}
\usepackage{bm}
\usepackage{braket}

\begin{document}

\preprint{APS/123-QED}

\title{Three-dimensional Micromotion Compensation Protocols for an RF Ion Trap}

\author{Ryoichi Saito$^{1}$}
\email{r-saito@phys.sci.isct.ac.jp}
\author{Takashi Mukaiyama$^{1}$}%
\email{mukaiyama@phys.sci.isct.ac.jp}
\affiliation{
$^1$Department of Physics, Institute of Science Tokyo, Ookayama 2-12-1, Meguro-ku, Tokyo 152-8550, Japan\\
}%

\date{\today}

\begin{abstract}
We propose and demonstrate four procedures for three-dimensional micromotion compensation by combining two methods: the RF-photon correlation method and the displacement method based on trap RF amplitude modulation.
In ion traps, the structure of the electrodes or the vacuum chamber may restrict the laser beam incidence direction, and causes the limitation in the compensation scheme.
We present four protocols, ensuring that at least one of them can be applied to experimental systems with various individual constraints.
We also discuss the compensation accuracy and practical applicability of each of these four approaches.
This work provides a practical guideline for performing full three-axis micromotion compensation and contributes to the advancement of endcap traps, which are highly suitable for single-ion trapping.
\end{abstract}

\maketitle


\section{\label{sec:level1}Introduction}
Laser-cooled trapped ions are one of the foundational technologies underpinning the rapidly advancing field of quantum technologies.
Ion traps have been developed and implemented in various electrode configuration designs to suit specific applications and the number of ions to be trapped.
In recent years, the development of quantum information science and quantum computing research has brought attention to conventional linear ion traps~\cite{Nagerl1998, Schmidt-Kaler2003}, which are effective at trapping multiple ions.
On the other hand, there also exist ion traps optimized for capturing a single ion. Endcap traps~\cite{SCHRAMA199332}, and stylus traps~\cite{Maiwald2009, 10.1063/1.4817304} are representative examples.
Such ion traps, which are highly effective for single-ion trapping, hold promise for applications in optical frequency standards~\cite{Gill_2005} and, in the case of stylus traps, sensing applications due to their superior accessibility.

Micromotion compensation is essential as a foundational technology for conducting quantum technology research and precision measurement research using ion traps. This technique minimizes the effects of the RF field used for ion trapping and is one of the critical methods for reducing ion heating.
Various methods have been developed over time, including the RF-photon correlation method~\cite{berkeland1998minimization}, parametric excitation~\cite{Narayanan, Tanaka}, atomic loss due to atom-ion collisions\cite{harter}, ion trajectory analysis~\cite{PhysRevA.92.043421}, RF amplitude modulation–induced ion displacement~\cite{10.1063/5.0046121}, Ramsey interferometry~\cite{higgins2021micromotion}, qubit transition and direct scanning of dc voltage~\cite{Lee:2}, dark and bright ion crystal~\cite{barnea2025}, dark resonance~\cite{dark, Allcock_2010}.

The selection of micromotion compensation methods and protocols is closely tied to trap geometry.
Regarding trap geometry, in the case of a linear ion trap, ions are confined in two dimension with RF fields, while confinement along the remaining axis is accomplished using an electrostatic field.
Therefore, ideally, compensation is only required along two axes.
On the other hand, in endcap ion traps and stylus traps, ions are confined in all three directions with RF fields, necessitating compensation in all three directions.
The widely used RF-photon correlation method is sensitive to micromotion in the direction of the cooling laser's incidence, requiring compensation to be performed by independently irradiated lasers from multiple directions.
Furthermore, when employing a method to minimize the displacement of the ion's trapping position caused by changes in RF voltage, positional changes are observed using fluorescence images, making it difficult to detect displacements in the vertical direction of the fluorescence image.
Therefore, it is necessary to measure positional changes in the depth direction by scanning the objective lens~\cite{PhysRevA.92.043421, 10.1063/5.0046121} or using super-resolution imaging techniques~\cite{suprereso}.
As described above, the required elements vary depending on trap geometry and method employed, making it essential to select appropriate methods and equipment and establish well-suited protocols.

In this study, we propose four practical protocols for three-axis micromotion compensation.
Even under experimental constraints that arise when constructing actual ion trap systems, it should be possible to find a suitable protocol tailored to each specific setup from the protocols presented in this paper.
The four protocols are based on two micromotion compensation techniques: the RF-photon correlation method and the displacement method using RF voltage modulation.
We demonstrate these protocols through experiments using an endcap trap, a representative ion trap configuration that confines ions along all three spatial axes with RF electric fields.
To evaluate the compensation accuracy of each protocol, we measured the residual electric fields and confirmed that all methods were capable of achieving compensation with an accuracy on the order of a few volts per meter.
Although endcap traps are widely used as a general-purpose ion trap configuration, there has been no prior study that systematically investigates their micromotion compensation procedures.
The outcomes of this study offer practical solutions for cases requiring full three-axis micromotion compensation, particularly in endcap traps, and contribute to the broader ion trap research community.

\begin{figure}[tb]
	\centering
	\includegraphics[width=7.5cm]{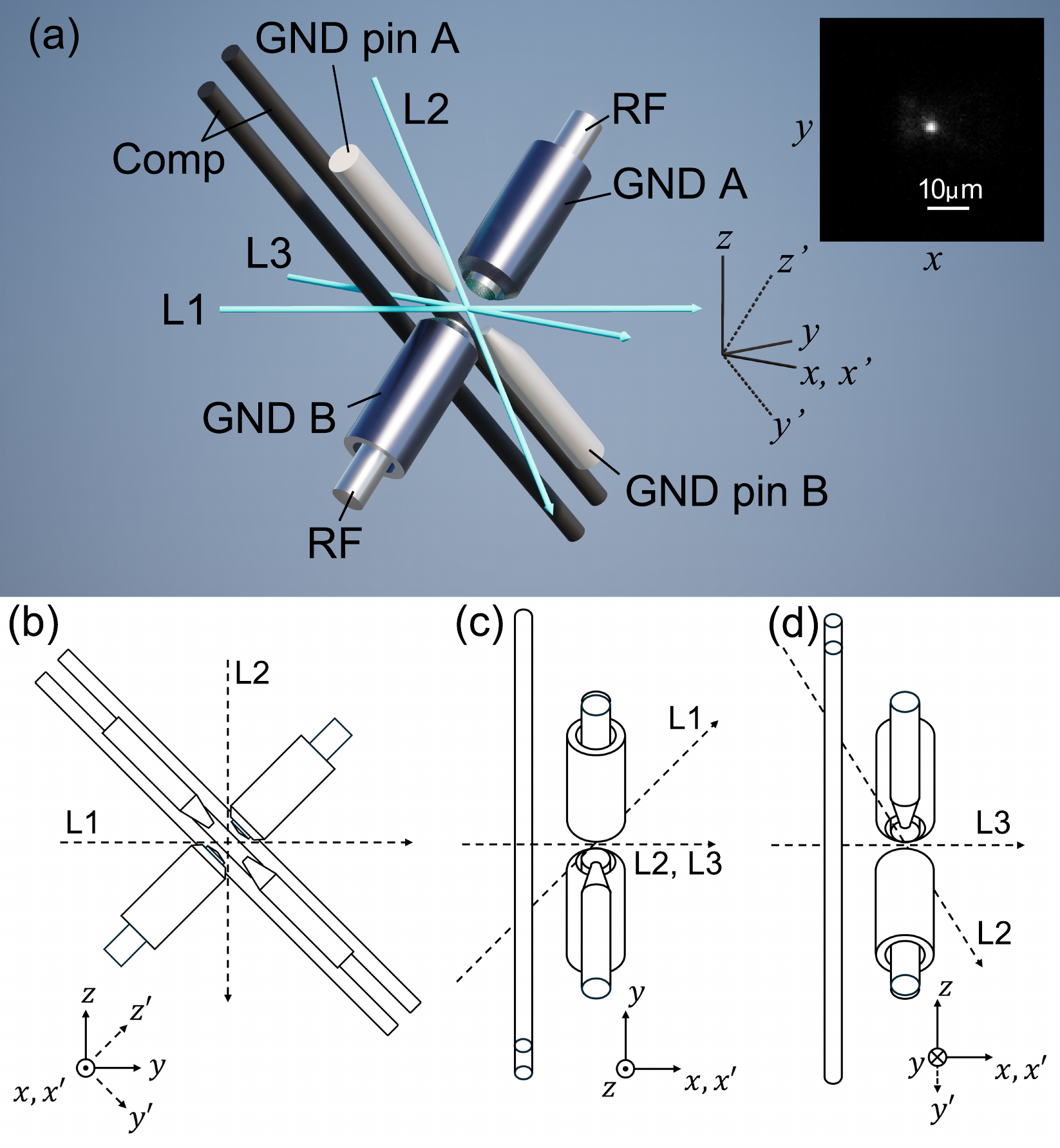}
	\caption{
        (a) Schematic diagram of the endcap trap used in this study.
        Cooling lasers are introduced from three directions, labeled L1, L2, and L3.
        The inset in the upper right shows an image of a single trapped ion.
        The camera's imaging plane corresponds to the $xy$-plane.
        (b)(c)(d) Schematic diagrams of the endcap trap viewed from the $x$-axis, $z$-axis, and $y$-axis directions, respectively.
        The cooling laser beams are indicated by dashed lines.
        }
	\label{fig:fig1}
\end{figure}


\section{\label{sec:level2}Experimental setup}
In this study, we utilized a conventional endcap trap, as shown in Fig.~\ref{fig:fig1}(a).
This ion trap consists of a pair of RF rods, a pair of cylindrical GND electrodes, a pair of GND pin electrodes, and two Comp rod electrodes.
An RF voltage with an amplitude of 565~V and a frequency of 25.7~MHz was applied to the RF rod electrodes, while the GND electrodes were grounded relative to the RF voltage.
As a result, the three trap principal axes are expected to form along the $z'$ axis, which connects the RF rod electrodes, and two orthogonal axes on the $x'y'$ plane.
Due to the presence of the GND pin electrodes, the RF field bends, and the trap principal axes are expected to form along axis which connects GND pin electrodes ($y'$ axis) and its perpendicular axis ($x'$ axis).
To prevent the degeneracy of the trap frequencies along the two axes on the $x'y'$ plane, an offset voltage of 0.4~V was applied to the GND pin electrodes.
As a result, the trap frequencies of each axes were found to be $(\omega_{x’}, \omega_{y’}, \omega_{z’}) = 2\pi \rm\times(198~kHz, 236~kHz, 440~kHz)$~\cite{PhysRevA.104.053114, 10.1063/5.0100007}.
The micromotion compensation in three axes was performed by applying voltages to the GND A, GND pin A, and Comp electrodes.
For the two Comp electrodes, the same voltage was applied to approximately compensate for the electric field in the $x'$ direction.
The GND pin A electrode primarily compensated for the electric field in the $y'$ direction, and the GND A electrode compensated for the electric field in the $z'$ direction.

The ion trap installed inside the vacuum chamber was oriented such that the $z'$ axis, which connects the RF electrodes, was tilted 45 degrees relative to the direction of gravity($z$ axis).
This configuration was chosen to enable fluorescence observation of the ions along the $z$ axis.
In this experiment, we captured $^{174}\mathrm{Yb}^+$ ions.
Yb atomic vapor was generated using a current-heated oven placed near the trap, and $^{174}\mathrm{Yb}^+$ ions were produced by photoionization using a 399~nm laser and a 369~nm laser, both of which were incident from the $y$ axis direction.
The trapped ions were Doppler-cooled using the 369~nm cooling laser beams incident in directions L1, L2, and L3, as shown in Fig.\ref{fig:fig1}(a)-(d), along with a 935~nm repump laser incident from the $y$ axis direction.
The fluorescence emitted by the trapped ions in the z-direction was collected by an objective lens, split by a beam splitter, and detected by both an EMCCD camera and a PMT.
Thus, the camera captured images in the $xy$ plane, and a typical fluorescence image of a single ion is shown in the inset.

Table~\ref{tb:table0} presents the list of three-dimensional compensation protocols proposed and demonstrated in this paper.
Three-dimensional micromotion compensation is performed by combining two different compensation methods: the RF-photon correlation method and the displacement method.
Since either of these two methods can be applied to each axis, we propose and demonstrate four protocols, designated Protocols A through D.
Details of the specific methods used in each protocol and the corresponding results are presented in \ref{sec:level3}.

\begin{table}[tbh]
\centering
\caption{List of protocols for three-dimensional compensation. }
\label{tb:table0}
 \begin{tabular}{lr}\hline
    Protocol&Compensation scheme\\ \hline
     A&RF-photon correlation in all 3D\\
     B&2D RF-photon correlation and 1D displacement\\
     C&1D RF-photon correlation and 2D displacement\\
     D&Displacement in all 3D\\
     \hline
 \end{tabular}
\end{table}

\section{\label{sec:level3}Result}
\subsection{\label{level1}Compensation by RF photon correlation in all 3D}
We compensated for three-dimensional micromotion using the RF-photon correlation method.
To achieve this, we measured the correlation between the timing of ion fluorescence emission and the phase of the trap RF signal.
Photon pulses detected by the PMT were used as start pulses, while rectangular pulses derived from the trap RF signal served as stop pulses. These signals were input into a Time-to-Amplitude Converter (TAC), and a typical dataset consisted of 2000 events to generate the correlation signal.

Cooling lasers were injected along three directions, labeled L1, L2, and L3, to search for the compensation position.
Each laser was designed to be incident from a different direction, such that they had components along the trap’s principal axes, characterized by $x’$, $y’$, and $z’$.
L1 was aligned in the $xy$ plane, entering at a 45~degree angle to both the $x$ and $y$ axes.
L2 propagated in the $zx$ plane, making an angle of 31~degrees with respect to the $z$ axis.
L3 was aligned parallel to the $x$ axis, passing through the trap center.

\begin{figure}[tb]
	\centering
	\includegraphics[width=6.8cm]{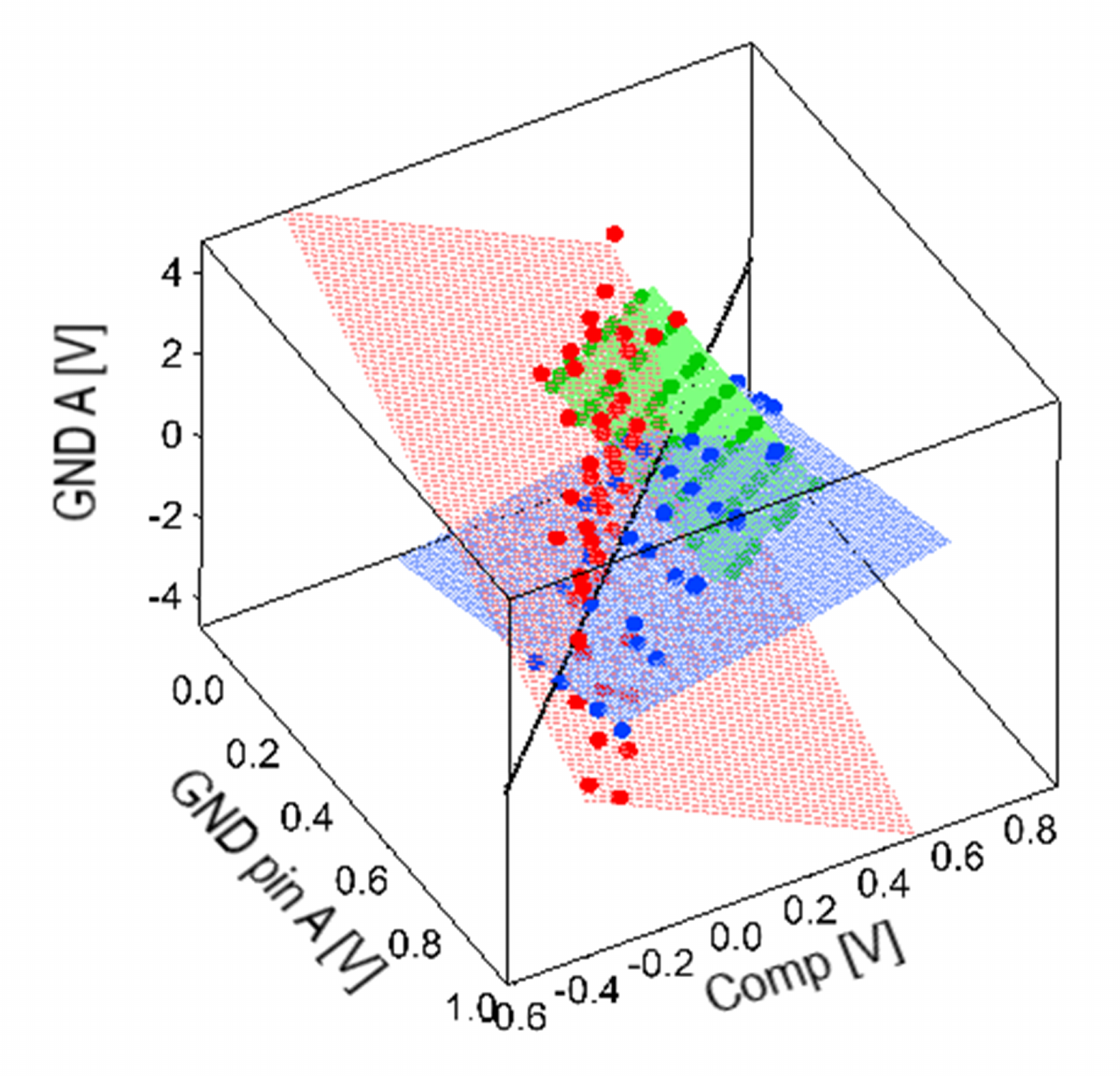}
	\caption{
     Compensation point obtained by the RF-photon correlation method in the compensation voltage space.
        The red, blue, and green data points correspond to results obtained using the cooling lasers L1, L2, and L3, respectively.
        Each colored plane represents the result of fitting a planar function to the measurement points for each laser.
        The solid black line indicates the line of intersection between the red and blue planes.
    }
	\label{fig:fig2}
\end{figure}

The method for identifying the compensation point is described below.
While one of the three cooling lasers was incident, we varied one of the three voltages applied to the electrodes, typically scanning the GND pin A voltage while fixing the GNDA and Comp voltages, and acquired the correlation signal.
The correlation signal exhibits a sinusoidal profile over one cycle of the trap RF. 
By evaluating the amplitude of this signal as a function of the GND pin A voltage allows us to identify the voltage value at which the amplitude is minimized. This point corresponds to a low correlation and can thus be considered the compensation point for the given pair of GND A and Comp voltages.
The results of plotting these points are shown in Fig.\ref{fig:fig2}.

By repeating this procedure for various combinations of GND A and Comp voltages, we obtained compensation points in three-dimensional voltage space as a plane.
Among the results, red filled circles indicate measurements taken using cooling laser L1, blue filled circles with L2, and green filled circles with L3.

We fitted the compensation points obtained from each laser direction with a plane function, and the results are shown in Fig.\ref{fig:fig2}.
The red, blue, and green planes represent the fitted planes for the data obtained using the L1, L2, and L3 cooling lasers, respectively.
The intersection point of these three planes corresponds to a position where the correlation is minimized from all three laser directions (L1, L2, and L3).
From this, we determined the compensation point to be (GND pin A, Comp, GND A) = (0.272V, 0.386V, -0.386V).

\subsection{\label{level2}Compensation by 2D RF-photon correlation and 1D displacement}
Next, we performed compensation by combining the RF-photon correlation method with trap displacement.
In the previously described method, three-dimensional compensation was achieved solely using the RF-photon correlation technique with cooling lasers incident from three different directions.
In contrast, in this protocol, we used the RF-photon correlation method with cooling lasers from only two directions to identify the compensation condition, and aimed to determine the full three-dimensional compensation point by utilizing displacement in the remaining direction.

An overview of the compensation protocol used in this study is presented below.
First, we irradiated the ion with cooling lasers L1 and L2, respectively, scanned the compensation voltages, and determined the compensation condition using the RF-photon correlation method.
As a result, we obtained the red and blue compensation points shown in Fig.\ref{fig:fig2}, as described earlier.
Each dataset was fitted with a plane function, resulting in the red and blue planes shown in the figure.
From these two planes, their line of intersection is obtained, as indicated by the solid black line in Fig.\ref{fig:fig2}.
Along this line, any compensation voltage values are indistinguishable using the RF-photon correlation method when observed from either the L1 or L2 laser directions.
In the previously discussed method using only the RF-photon correlation technique, a third cooling laser (L3) was used to identify a unique point along this line and thereby determine the full three-dimensional compensation point.
In contrast, in the present method, we successfully determined the three-dimensional compensation condition by evaluating the displacement of the trap position along the intersection line when modulating the RF voltage, without using the third cooling laser.

In the displacement measurement, the amplitude of the trap RF was typically reduced from 565 V to 416 V in this study, and the ion's trapping position before and after the change in RF amplitude was observed via fluorescence imaging to search for the compensation condition.
If the stray electric field is not properly canceled, weakening the confinement causes the ion's trapping position to shift significantly due to the influence of the stray field. However, when the confinement is strong, the influence of the stray field becomes relatively small, and the ion's trapping position approaches the position corresponding to the compensated micromotion.
By utilizing this principle, we searched along the intersection line for the point where the ion's displacement before and after the RF modulation was minimized.

\begin{figure}[tb]
	\centering
	\includegraphics[width=8.5cm]{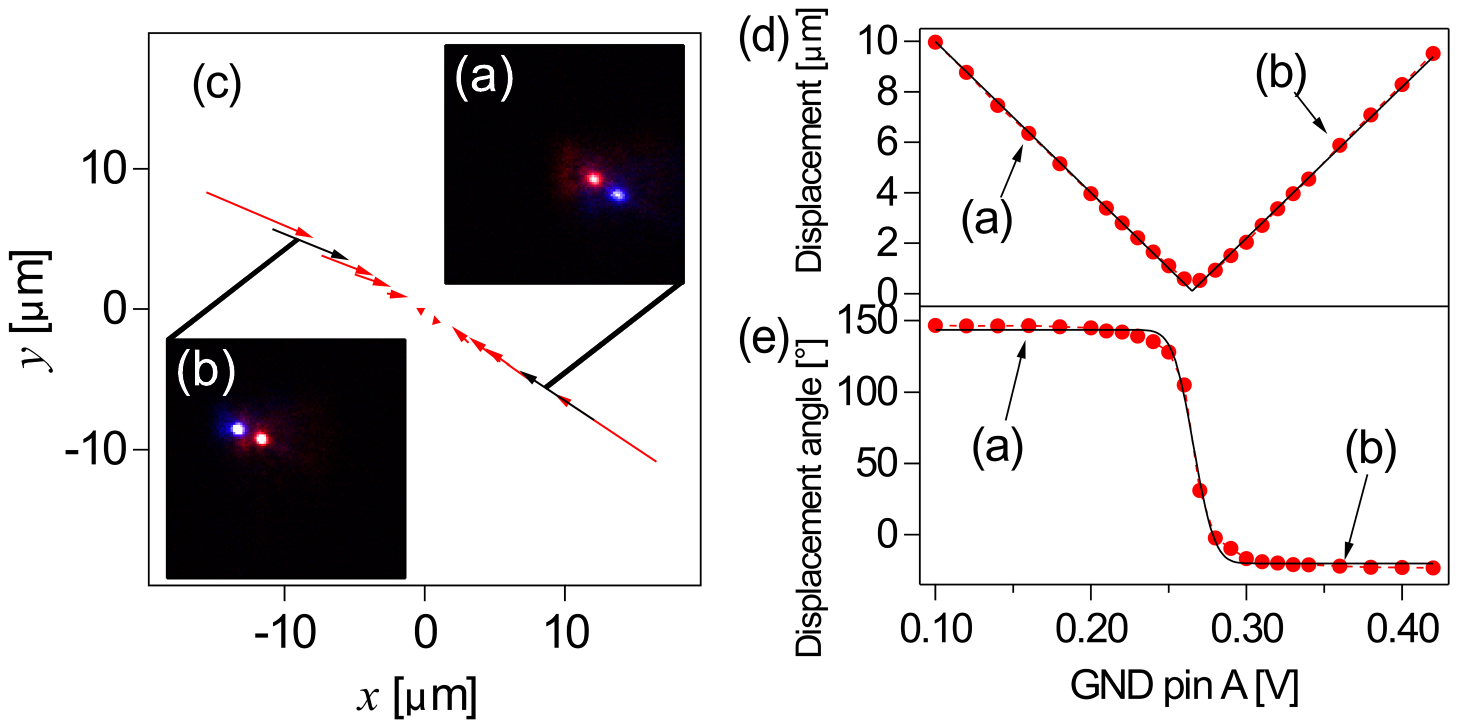}
	\caption{
    (a), (b) Fluorescence images of the ion before and after modulation of the trap RF amplitude.
    The blue bright spot indicates the ion position when the RF amplitude is low, while the red bright spot represents the ion capture position when the RF amplitude is high.
    (c) Displacement results along the intersection line of the two planes determined by the RF-photon correlation method.
    The vertical and horizontal axes represent the real-space displacement converted from camera coordinates using the imaging magnification.
    Each arrow corresponds to a displacement measurement at a different compensation voltage along the intersection line.
    The starting point of each arrow represents the ion position at low RF amplitude, and the end point represents the ion position at high RF amplitude
    (d) Variation of displacement magnitude along the intersection line.
    The horizontal axis represents the voltage value of GND pin A as a representative parameter.
    The solid black line shows the result of fitting with a V-shaped function.
    (e) Variation of the displacement direction angle along the intersection line.
    The horizontal axis again represents the voltage value of GND pin A.
    The solid black line indicates the result of fitting with an error function.
    }
	\label{fig:fig3}
\end{figure}

Figure~\ref{fig:fig3}(a) shows a typical experimental result of RF amplitude modulation at the point (GND pin A, Comp, GND A) = (0.160 V, 0.535 V, -0.437 V) along the intersection line.
The blue fluorescent spot corresponds to the ion image under weak confinement, while the red spot corresponds to that under strong confinement.
It can be seen that the ion’s trapping position is shifted due to the influence of the stray electric field.
To make it easier to visualize, the ion positions on the camera under weak and strong confinement are indicated as the start and end points of arrows in Fig.~\ref{fig:fig3}(c).

In contrast, Fig.~\ref{fig:fig3}(b) shows the result of the displacement measurement at a different point along the intersection line, (GND pin A, Comp, GND A) = (0.360 V, 0.269 V, -0.344 V).
Compared to Fig.~\ref{fig:fig3}(a), the direction of the change in the ion's trapping position is reversed.

Figure~\ref{fig:fig3}(c) shows a subset of the displacement measurements along the intersection line.
The length of the arrows decreases as they approach the origin on the graph.
Additionally, the direction of the arrows reverses across the origin.
This clearly indicates that as the point approaches the origin, the stray electric field is increasingly canceled out, resulting in reduced displacement.
When the point moves beyond the compensation point, the direction of the stray field reverses, leading to a reversal in the direction of the displacement.

Figures~\ref{fig:fig3}(d) and (e), respectively, show the plots of the displacement magnitude and the angle of the arrows shown in Fig.~\ref{fig:fig3}(c), both measured along the intersection line.
Although the combinations of GND pin A, Comp, and GND A voltages were varied to perform the measurements along the intersection line, the horizontal axis in Fig.~\ref{fig:fig3}(d) and (e) represents the GND pin A voltage as a representative parameter.

To evaluate the compensation point, Fig.~\ref{fig:fig3}(d) and (e) were fitted using a V-shaped function and an error function, respectively, and the results are shown as solid black lines in the figures.
The point where the displacement is minimized and the point where the displacement angle reverses are in good agreement, indicating that this position corresponds to the three-dimensional compensation point.
As the final result, the compensation point was determined to be (GND pin A, Comp, GND A) = (0.265 V, 0.396 V, -0.388 V).
As demonstrated above, three-dimensional compensation was successfully achieved by applying the RF photon correlation method in two directions and using displacement measurements for the remaining one-dimensional direction.

\subsection{\label{level3}Compensation by 1D RF-photon correlation and 2D displacement}
Here, we introduce a method in which the RF-photon correlation technique is applied using only a single cooling laser direction, while displacement measurements are used to determine compensation in the remaining two directions.
The protocol for this method begins by using the RF-photon correlation technique with one direction of cooling laser to obtain a plane of compensation, as described earlier.
By scanning this plane and evaluating the displacement magnitude, the three-dimensional compensation point is identified.

In this demonstration, the cooling laser L1 was used to perform RF-photon correlation measurements while scanning the compensation voltages, and a point with minimal correlation was found.
As previously described, this corresponds to the red data points obtained using RF-photon correlation with L1 in Fig.~\ref{fig:fig2}.
On the plane fitted to these measurement points, we scanned the voltages and measured the displacement.

In the displacement measurement, the trap RF amplitude was modulated, and the ion's trapping position before and after the modulation was measured using fluorescence imaging.
Fluorescence images were taken at RF amplitudes of 565 V and 416 V, and the distance between the ion's trapping positions at these two settings was evaluated.
Note that the fluorescence images represent projections onto the $xy$ plane.

\begin{figure}[tb]
	\centering
	\includegraphics[width=8.5cm]{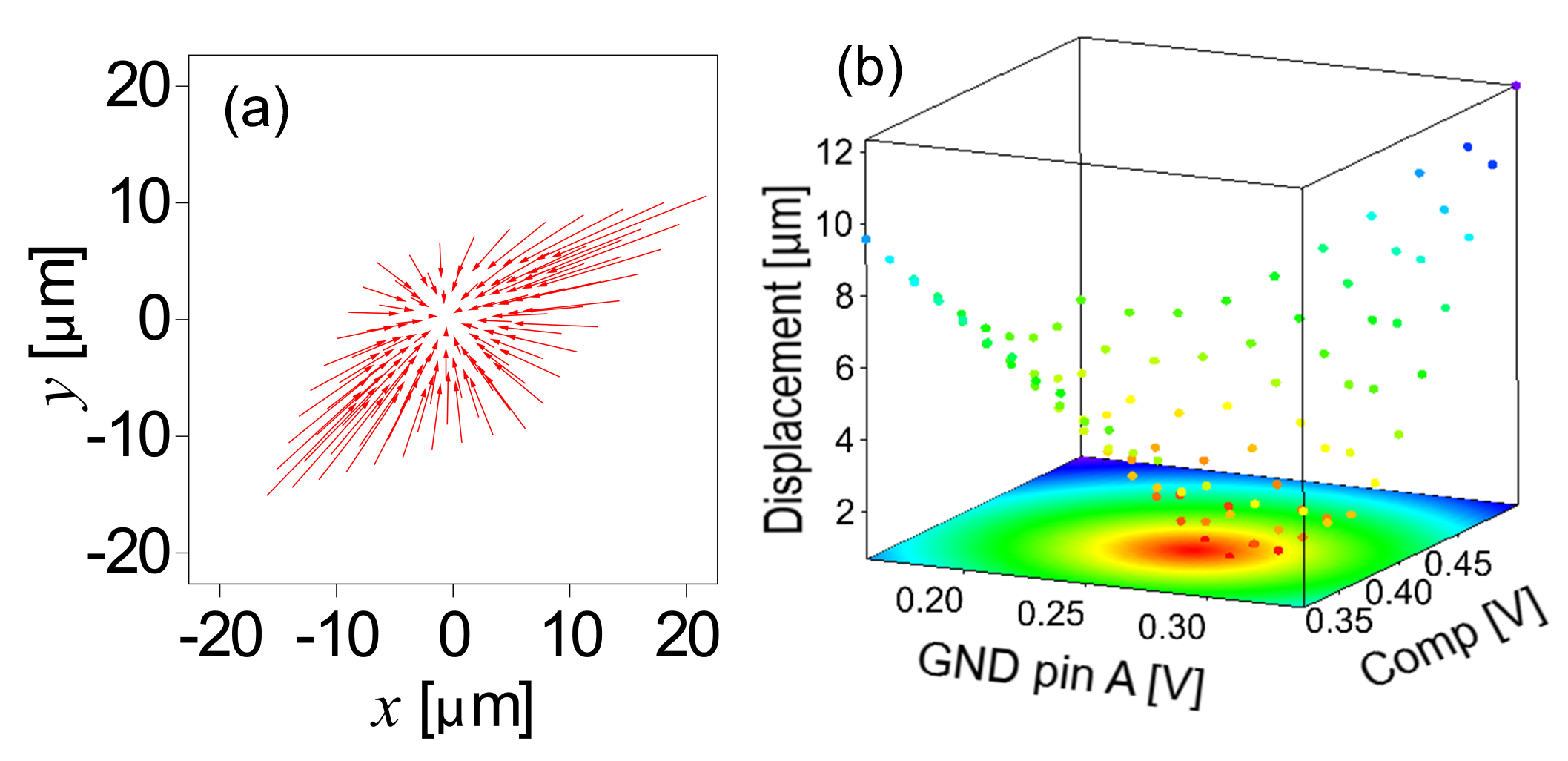}
	\caption{
    (a) Displacement measurement results on the compensation voltage plane obtained using the RF-photon correlation method.
    Each arrow represents the displacement on the camera corresponding to RF modulation at a given set of compensation voltages on the plane.
    The starting point and end point of each arrow indicate the ion position on the camera when the confinement is weak and strong, respectively.
    (b) Displacement magnitude on the compensation voltage plane.
    The vertical axis shows the length of the arrows in (a), representing the displacement size.
    The compensation voltage plane is represented using the GND pin A and Comp voltages.
    The color plot on the base represents the fitting result using a conical function.
    }
	\label{fig:fig4}
\end{figure}

Figure~\ref{fig:fig4}(a) shows the change in ion positions during displacement.
Each arrow represents a different combination of compensation voltages on the plane (GND pin A, Comp, GND A), obtained using the RF-photon correlation method.
In practice, the voltages on GND pin A and Comp were varied on a 0.02~V grid, with measurements taken at 100 points in total.
The starting point of each arrow indicates the ion position when the trap RF amplitude is low, while the end point indicates the position when the RF amplitude is high.
The imaging plane of the camera was converted based on the imaging magnification.
It is observed that when the RF amplitude is low and the confinement is weak, the ion is captured farther from the compensation point due to the influence of stray electric fields.
In contrast, when the confinement is stronger, the ion approaches the compensation point.

In the displacement results shown in \ref{level2}, the ion positions were scanned along a line determined by the RF-photon correlation method using two cooling beams to locate the compensation point.
In contrast, in the present case, the scan is performed over a plane determined by the RF-photon correlation method using a single cooling beam, resulting in a two-dimensional distribution of the arrows.
It is expected that the ion capture position at the compensation point lies at the center point toward which the arrows are directed.
Furthermore, it can be observed that the arrows become shorter as they approach the compensation point.
This trend is consistent with the situation in the method described in \ref{level2}, and occurs because the stray electric fields are being canceled out, allowing the ion to approach the compensation point.

To quantitatively assess the compensation point, the displacement magnitude represented by the length of each arrow, was plotted as a function of the compensation voltages within the plane determined by the RF-photon correlation method. This plot is shown in Fig.~\ref{fig:fig4}(b). Although all three compensation voltages were varied during the scan across the compensation plane, the figure displays the results using GND pin A and Comp voltages as representative parameters.

We fitted the results using a conical function to determine the optimal compensation point.
As a result, the compensation voltages were obtained as (GND pin A, Comp, GND A) = (0.273~V, 0.391~V, -0.409~V).

This method closely resembles the compensation method we previously presented for a linear trap.
In the earlier approach, the linear trap configuration allowed us to ignore axial micromotion and focus solely on compensating micromotion in the radial plane using displacement measurements.
In contrast, the present method differs in that it first identifies the compensation plane using the RF-photon correlation method and performs compensation along a single initial axis.
Subsequently, by scanning within the compensation plane and evaluating the displacement, compensation along the remaining two axes is carried out.

\subsection{\label{level4}Compensation by displacement in all 3D}
The results of micromotion compensation along all three axes using displacement measurements are presented.
In this demonstration, displacement was measured at 512 points by scanning the compensation voltages of GND pin A, Comp, and GND A over an 8-point grid with 0.05~V steps for each voltage.

\begin{figure}[tb]
	\centering
	\includegraphics[width=8cm]{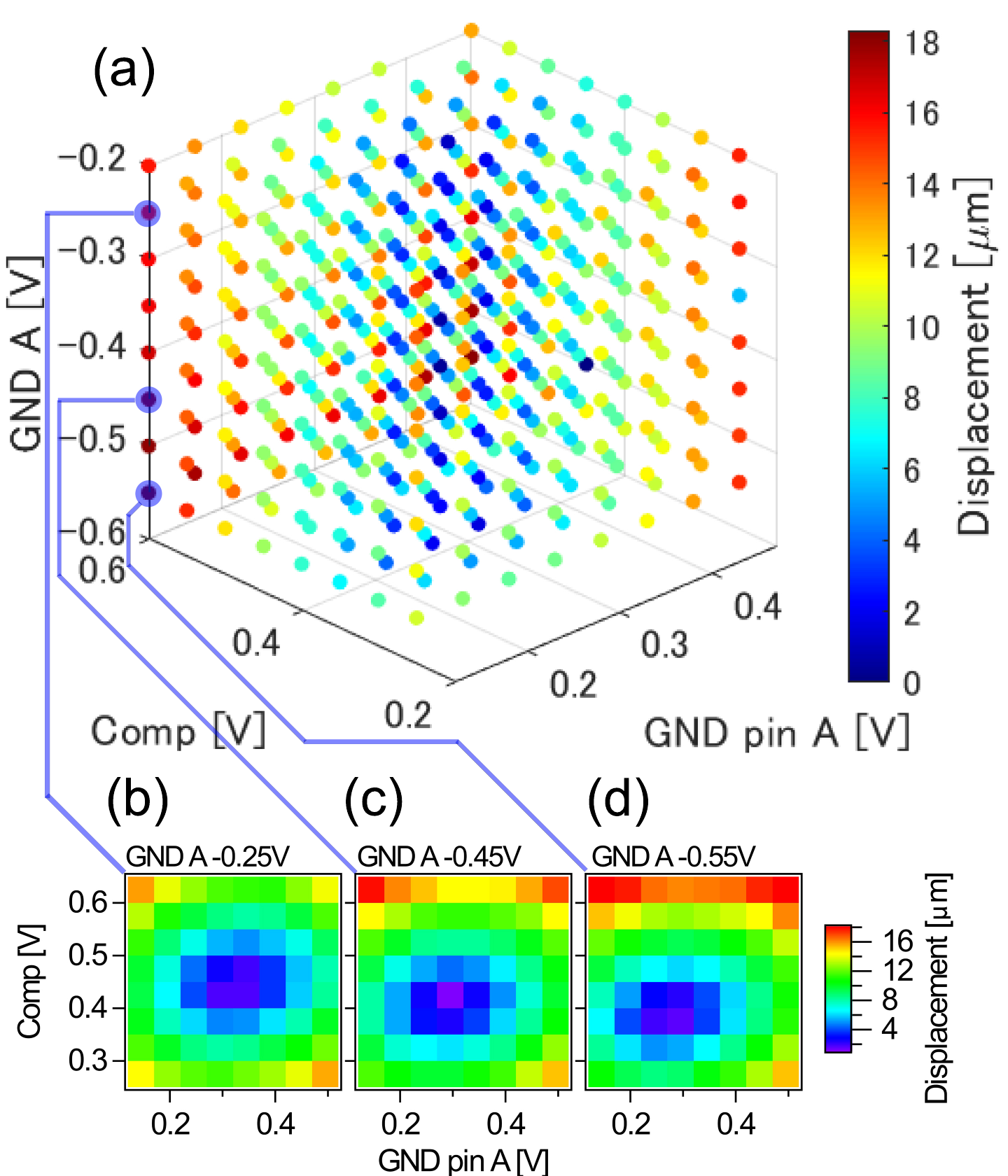}
	\caption{
    (a) Displacement size measurement results in three-dimensional compensation voltage space.
    The three axes represent the voltages applied to the three compensation electrodes, and the displacement size at each point is shown as a color plot.
    (b)(c)(d) Sliced views of the results in (a) at GND A voltage values of -0.25~V, -0.45~V, and -0.55~V, respectively.
    The horizontal and vertical axes correspond to the Comp and GND pin A voltages, and the displacement size is represented as a color plot.
    }
	\label{fig:fig5}
\end{figure}

In the displacement measurements, the distance between the ion position on the camera at low trap RF amplitude and its position at high trap RF amplitude was evaluated.
Figure~\ref{fig:fig5}(a) shows the result of plotting these position changes on the camera for each set of compensation voltages.
The displacement size was converted into actual physical units based on the camera’s imaging magnification, and is displayed as a color map in the figure.

Figures~\ref{fig:fig5}(b), (c), and (d) show cross-sectional views of the graph in (a) at specific values of the GND A voltage.
From these displacement cross-sections, it can be seen that compensation voltages yielding small displacement sizes form a conical shape.
Furthermore, examining the minimum displacement sizes across different cross-sections reveals that there is also an optimal voltage along the GND A direction.
Therefore, by fitting the data with a four-dimensional conical function, the compensation voltages were determined to be (GND pin A, Comp, GND A) = (0.276~V, 0.393~V, -0.422~V).

\subsection{\label{level5}Disccusion on the Performance}
In this section, we evaluate and compare the performance of each protocol, and further examine them in terms of the accuracy of micromotion compensation, simplicity of the protocol, and ease of implementation.

In this demonstration, we present the results for four types of protocols: one using the RF-photon correlation method, two combining the RF-photon correlation method with trap RF amplitude modulation-induced displacement, and one using displacement alone.

\begin{table}[tbh]
\centering
\caption{Compensation voltages and residual electric fields obtained for each micromotion compensation protocol.
The voltage values applied to the compensation electrodes—GND pin A, Comp, and GND A—obtained for each protocol described in \ref{sec:level3}.
Additionally, the estimated residual electric field for each protocol was evaluated individually for each compensation electrode.}
\label{tb:table1}
 \begin{tabular}{lrrrr}\hline
     Protocol& A& B& C& D \\ \hline
     GND pin A& 0.272~V& 0.265~V & 0.273~V & 0.276~V  \\
     Comp & 0.386~V & 0.396~V & 0.391~V & 0.393~V \\
     GND A & -0.386~V & -0.388~V & -0.409~V & -0.422~V \\ \hline
     $\Delta E_{\rm GND pin A}$ & 0.177~V/m & 0.620~V/m & 1.61~V/m & 2.28~V/m \\
     $\Delta E_{\rm Comp}$ & 0.979~V/m & 0.399~V/m & 1.04~V/m & 4.78~V/m \\
     $\Delta E_{\rm GND A}$ & 2.33~V/m & 1.950~V/m & 5.07~V/m & 18.3~V/m \\ \hline
 \end{tabular}
\end{table}

Table \ref{tb:table1} shows the compensation voltages obtained using the four protocols.
These are the voltages applied to the compensation electrodes: GND pin A, Comp, and GND A.
In this demonstration, the four protocols were carried out consecutively within a short period to minimize temporal variations such as changes in stray electric fields.
The experiments were conducted in the order of protocols A, B, C, and D.
If temporal variations can be neglected, all compensation voltages should ideally be identical.
The measured voltages are generally consistent, with the maximum relative deviations for GND pin A, Comp, and GND A being 4\%, 2.5\%, and 8.8\%, respectively.
Notably, for GND A, which exhibited the largest deviation, the voltage decreased sequentially from A to D, suggesting that there may have been a time-dependent change requiring compensation in the direction compensated by GND A.

Next, in order to evaluate the accuracy of the compensation, the residual electric field was estimated.
In the table, $\Delta E_{\rm GND\ pin\ A}$, $\Delta E_{\rm Comp}$, $\Delta E_{\rm GND A}$ represents the residual electric fields in the directions of GND pin A, Comp, and GND A electrodes, respectively.
Micromotion has been compensated at a level comparable to the residual electric fields reported in previously proposed compensation methods.

The results indicate that the accuracy improves in the order of Protocol D, C, B, and A.
However, the accuracy of the compensation using displacement can be improved by increasing the resolution of the compensation voltage steps near the compensation point and by increasing the modulation of the trap RF amplitude.
For these reasons, the compensation accuracy for Protocols B, C, and D can likely be further improved.

Since the RF-photon correlation method is sensitive to micromotion along the direction of the cooling laser, lasers must be injected from three directions in order to compensate micromotion along all three axes.
Therefore, in this study, we introduced lasers L1, L2, and L3, and used the RF-photon correlation method with each to search for the compensation point.
The trap and vacuum chamber must be structured to allow cooling light to be introduced from three directions.
This is an important consideration that must be carefully addressed during the design phase.

In the case of displacement measurements, it is sufficient to perform imaging; therefore, the direction of laser incidence does not require particular consideration.
To observe three-dimensional positional shifts induced by amplitude modulation as two-dimensional displacements on the camera plane, it is necessary to adjust the lens position or employ other methods to extract positional information along the depth axis.
Conversely, if depth measurements are not performed, it is advisable that the fluorescence observation axis be intentionally misaligned with the principal axis of the trap.

\section{\label{sec:level4}Conclusion}
In this study, we investigated the protocol for compensating three-dimensional micromotion in an endcap trap.
To compensate for micromotion along all three axes, it is necessary to cancel out the stray electric fields in three-dimensional space.
Therefore, we prepared three compensation electrodes to enable the application of electric fields for micromotion compensation

We developed compensation protocols by combining two micromotion compensation methods: the RF photon correlation method and displacement measurement based on trap RF amplitude modulation.
Using this approach, we established a total of four three-dimensional micromotion compensation protocols.
We demonstrated each of these protocols experimentally and showed that an optimal compensation point could be obtained for all of them.

The compensation voltages obtained from the demonstration of the developed protocols were within a few percent relative error, indicating that each method successfully identified the correct compensation point.
To evaluate the accuracy of the compensation procedures, the residual electric fields were assessed for each protocol and each compensation electrode.
The residual electric fields observed in this demonstration were found to be comparable to those reported in previously proposed and demonstrated compensation methods.
Furthermore, we discussed the ease of implementation and the specific requirements associated with each compensation method, providing information that enables the selection of an appropriate protocol according to the experimental objective.

As described above, we have systematically presented a set of compensation protocols and their performance for a versatile three-dimensional micromotion compensation method in endcap traps.
These results represent a significant contribution to the future development and advancement of endcap trap systems.

\section*{\label{sec:level5}Acknowledgments}
This work is supported by Co-creation Place Formation Support Program Grant Number JPMJPF2015, and JSPS KAKENHI Grant Number JP23K13046.

\section*{\label{sec:level6}Data Availability}
The data that support the findings of this study are available
from the corresponding author upon reasonable request.



\end{document}